# Imaging Surface Magnetization in Altermagnetic MnTe Films

Ling-Jie Zhou[1], Senlei Li[1], Zi-Jie Yan[2], Yufei Zhao[2,3], Hongtao Rong[2], Zelong Xiong[1], Yiran Zhao[1], Zhaorong Gu[1], Pu Xiao[2], Ke Wang[4], Lok Kan Lai[2], Hyeonhu Bae[2], Haoyu Liu[1], Chao-Xing Liu[2], Binghai Yan[2,3], Cui-Zu Chang[2,*], Hailong Wang[1,*], and Chunhui Rita Du[1,*]

[1]School of Physics, Georgia Institute of Technology, Atlanta, GA 30332, USA
[2]Department of Physics, The Pennsylvania State University, University Park, PA 16802, USA
[3]Department of Condensed Matter Physics, Weizmann Institute of Science, Rehovot 7610001, Israel
[4]Materials Research Institute, The Pennsylvania State University, University Park, PA 16802, USA

[*]Corresponding authors: cxc955@psu.edu, hwang3021@gatech.edu, and cdu71@gatech.edu

**Abstract:** Altermagnets with pronounced spin-splitting band structure, unconventional magnetic and crystal symmetries, and exotic magneto-transport properties have received immense interest in cutting-edge spintronics, materials science, and condensed matter physics research. Microscopic imaging of spontaneous magnetic domains and phases in altermagnets constitutes an important step for investigating their underlying material properties, mechanisms, and spin behaviors. Taking advantage of scanning-probe quantum microscopy, here we report nanoscale quantum sensing of a prototypical altermagnet candidate $\alpha$-MnTe. We visualize evanescent magnetization and the associated magnetic domains in epitaxial MnTe films, which allows external magnetic fields to control the intrinsic altermagnetic order and configurations. By evaluating a series of MnTe films with different thicknesses down to the atomic scale, we further present evidence for the interfacial origin of the observed weak magnetization and show its correlation with the anomalous Hall effect in MnTe films. Our results advance the current understanding of emergent altermagnetism, providing insights into future material design of altermagnet-integrated spintronic devices.

## I. Introduction

The last couple of years witnessed the birth of "altermagnetism", a newly discovered magnetic phase at the forefront of quantum materials and spintronics research [1-8]. In contrast with their ferromagnetic and antiferromagnetic counterparts, altermagnets feature unconventional magnetic and crystal symmetries that enable nonrelativistic anisotropic spin-band splitting [4-12], anomalous Hall (AH) effect [13-15], tunable spin-current effects [16-18], and many other exotic magnetic/spin behaviors[19-22], providing a novel material platform for implementing advanced spintronics innovations. To date, a variety of altermagnet candidates have been explored in this context, and transformative spintronic concepts and devices are underway. Some of the notable examples in this catalog include spin-splitting torques [16-18], altermagnetic tunnel junctions [23,24], altermagnetization switching/control [25-27], and piezomagnetism [28,29].

Despite the enormous technological promise and rich fundamental physics, the ongoing "altermagnetism" research remains in its infancy. One of the major technical challenges results from the sparse experimental approaches capable of evaluating nanoscale spin properties of the nearly compensated altermagnetic order, which further impedes material design, performance testing and improvements of altermagnet-integrated spintronic systems. In the current state of the art, microscopic altermagnetic states/properties are typically inferred indirectly from magneto-transport measurements [14,15,25,30-34], which are susceptible to local thermal, electromigration, and magnetoelastic artifacts. While large-scale spectroscopies such as polarized X-ray photoemission electron microscopy have been utilized to investigate altermagnetic order and domains, they require a large-scale, high-energy synchrotron facility that is not widely available [19,20,35]. There is ongoing intense activity to develop and understand the family of altermagnetic materials, as well as to create new ones [36-40]. The success of these efforts relies simultaneously on advances in theory, material synthesis, and development of new experimental approaches.

Here, we introduce scanning-probe nitrogen-vacancy (NV) microscopy [41-50], a tabletop, advanced quantum imaging tool (see Supplemental Material Section I.1 [51] for details), to contribute to the rapidly growing altermagnetism research. Our work focuses on a representative altermagnet $\alpha$-MnTe [6,13,19]. Among the current altermagnet candidates, MnTe naturally stands out due to its high *Néel* temperature ($T_N$ ~310 K) [52-56], semiconducting nature [13,57,58], robust AH response [13,31-34,59-62], and large anisotropic spin-splitting that has been experimentally revealed by angle-resolved photoemission spectroscopy (ARPES) measurements [6,9,10,63,64]. The observation of the AH effect [13,31-34,59-62] and altermagnetic domain tunability under an external magnetic field [19] point to the presence of a net magnetization. Prior studies attribute the weak ferromagnetic response and AH effect in MnTe to different origins, such as intrinsic spin canting [31], interfacial conduction and ferromagnetism [33], and stoichiometry- or defect-induced ferromagnetism [34]. However, direct nanoscale imaging of the net magnetization in MnTe remains a formidable challenge, and its microscopic origin remains to be explored. Taking advantage of the excellent field sensitivity of NV sensors, we show that epitaxial MnTe films host a weak, yet finite out-of-plane (OOP) magnetization $M_z$, allowing external magnetic fields to control the domain configuration that is intrinsically correlated with the altermagnetic order. The weak net magnetization manifests a monotonic decrease with increasing temperature and coincides with the emergence of the AH effect in MnTe films. By investigating a series of MnTe samples with different thicknesses down to the atomic scale, we present evidence that the net magnetization arises predominantly from the sample surface, which is captured well by first-principles calculations of layer-resolved magnetic moments carried by Mn and Te atoms. Our results reveal the presence of evanescent magnetization in altermagnetic materials, providing insights into future studies of magnetic domain dynamics and domain manipulation in altermagnetic spintronic systems.

## II. Results

### A. Material platform: altermagnet α-MnTe

We first briefly review the pertinent material properties of altermagnet MnTe. Microscopically, MnTe has the NiAs crystal structure with collinear antiferromagnetic order (Fig. 1a). The magnetic moment carried by Mn atoms is parallel within the $c$-plane and antiparallel between two adjacent layers along the $c$-axis. The different arrangements of Te atoms around two neighboring Mn sublattices break the inversion-time reversal and translation-time reversal symmetry but preserve a screw rotation-time reversal symmetry [2,3,6,13]. The *Néel* vector, $\boldsymbol{L} = \boldsymbol{M}_1 - \boldsymbol{M}_2$, where $\boldsymbol{M}_1$ and $\boldsymbol{M}_2$ represent the magnetic moment of two antiferromagnetic spin sublattices, has six equivalent in-plane easy axes along the $[1\bar{1}00]$ crystallographic direction (Fig. 1b).

Next, we discuss the origin of the weak OOP magnetization formed in MnTe. MnTe possesses a mirror symmetry with respect to its basal plane (0001) [62]. Under a mirror operation, a pseudo vector transforms such that its components parallel to the mirror plane are reversed, while the component normal to the mirror plane remains invariant. The net magnetization of MnTe, $\boldsymbol{M} = \boldsymbol{M}_1 + \boldsymbol{M}_2$, is a pseudo vector. Consequently, its in-plane components are forbidden by the mirror symmetry, and the OOP component is allowed. As a result, an uncompensated net magnetization, if exists in MnTe, is expected to spontaneously orient along the $c$-axis direction. The symmetry argument presented above is in qualitative agreement with the recent theoretical work that predicts a weak, nonzero OOP magnetization $M_z$ induced by the interplay between spin-orbit coupling and altermagnetic order in MnTe [65]. The space group of MnTe forbids a linear coupling of $\boldsymbol{L}$ to $\boldsymbol{M}$, thus ruling out the usual bilinear Dzyaloshinskii–Moriya interaction (DMI). Instead, symmetry analysis reveals that the coupling between $\boldsymbol{L}$ and $\boldsymbol{M}$ emerges first at the third order of $\boldsymbol{L}$, taking the form $(3L_x^2 - L_y^2)L_y M_z = sin(3\phi_L)L^3 M_z$ [65,66]. Here, the $x$-axis and $y$-axis are along the $[2\bar{1}\bar{1}0]$ and $[1\bar{1}00]$ crystallographic directions, respectively, and $\phi_L$ is the azimuthal angle between the *Néel* vector $\boldsymbol{L}$ and the $[2\bar{1}\bar{1}0]$ axis of MnTe (Fig. 1b). Therefore, the theoretically predicted $M_z$ exhibits a characteristic threefold angular dependence $M_z \propto sin(3\phi_L)$ (Fig. 1c). The three magnetic easy axes with $\phi_L$ = 30°, 150° and 270° of the *Néel* order correspond to the upward $M_z$ while the remaining three states $\phi_L$ = 90°, 210° and 330° favor the downward $M_z$. The evanescent OOP magnetization of MnTe is estimated on the order of $10^{-5}$ $\mu_B$/Mn, which lies at the boundary or beyond the detection limit of conventional magnetometry techniques (see Supplemental Material Section I.2 [51] for details) [59,61]. To our knowledge, experimental observation of $M_z$ in altermagnetic MnTe films remains elusive [13,30,33,34,56].

### B. Scanning-probe quantum imaging of surface magnetization in MnTe films

In the current work, we utilize an NV center as a noninvasive probe to spatially image the weak uncompensated net magnetization in epitaxial MnTe films (Fig. 1d). Scanning-probe NV magnetometry exploits the linear Zeeman effect to detect a local magnetic stray field $B_{NV}$ longitudinal to the NV spin axis [41-44,46,48,67,68]. The magnitude of $B_{NV}$ is deduced from splitting of NV spin energies, which can be read out by optically detected magnetic resonance (ODMR) measurements [41,68] (see Supplemental Material Section I.3 [51] for details). The spatial resolution of scanning-probe NV magnetometry is solely determined by the NV-to-sample distance beyond the optical diffraction limit [67], which is ~100 nm in our measurements. Epitaxial (0001)-oriented MnTe films are grown on InP(111)A or GaAs(B) substrates by molecular beam epitaxy (MBE). A 10-nm-thick tellurium (Te) or 2-nm-thick bismuth (Bi) layer is capped at room temperature after the sample growth to prevent oxidation and other environmental effects. Prepared MnTe thin films are subsequently fabricated into micrometer-sized patterns using photolithography and Ar plasma etching methods (see Supplemental Material Section I.1 [51] for details). A cross-sectional scanning transmission electron microscopy (STEM) image presented in Fig. 1e reveals an atomically sharp interface between InP substrate and the MnTe film, which is free of stacking fault or local dislocations. The excellent crystalline quality of MnTe samples is further confirmed by the sharp and streaky reflection high-energy electron diffraction (RHEED) patterns measured throughout the growth processes and X-ray diffraction scan (see Supplemental Material Section I.2 [51] for details). We also perform ARPES measurements on an 8-unit-cell (UC) thick MnTe film to evaluate its band structure. The band dispersion along $\overline{\mathrm{K}} - \overline{\Gamma} - \overline{\mathrm{K}}$ is shown in Fig. 1f, where two surface bands ($\alpha_1$ and $\alpha_2$) and three bulk bands ($A_1$, B, and C) are resolved. The second derivative image presented in Fig. 1g highlights a pronounced spin splitting $\Delta \sim 280$ meV between the $A_1$ and $A_2$ bands, presenting evidence for altermagnetism formed in our epitaxial MnTe films [6,64,69]. Note that bands around the Fermi surface ($\alpha_1$ and $\alpha_2$) are attributed to the MnTe surface bands [62] (see Supplemental Material Section I.2 [51] for details).

We now report NV sensing of the MnTe films. Our measurements start with a hexagonal-shaped, 40-UC-thick MnTe pattern (with Te capping) prepared on an InP(111)A substrate (see Supplemental Material Section I.2 [51] for details). Figures 2a-2c show the maps for reconstructed magnetic stray field $B_z$ along $z$-axis measured at 2 K under three distinct field cooling processes from the room temperature: zero-field cooling with $B_{FC} = 0$ T, negative and positive field cooling with $B_{FC} = -1$ T and $+1$ T. The details of reconstructing $B_z$ from $B_{NV}$ can be found in Supplemental Material Section I.4 [51]. The magnetic cooling field $B_{FC}$ is applied along the OOP direction of the sample and is removed before subsequent NV measurements. Notably, robust magnetic stray field patterns are observed within physical boundaries of the sample, indicating uncompensated net magnetization hosted by the MnTe film. The zero-field cooling case ($B_{FC} = 0$) features a clear multidomain signature (Fig. 2a). Invoking the theoretical picture discussed above, the *Néel* vector $\boldsymbol{L}$ of MnTe is expected to be randomly distributed among the six equivalent easy axes in the $c$-plane without external field training processes, resulting in a largely equal domain population of the upward and downward OOP magnetization $M_z$ (Fig. 2d). Consistent with this expectation, our measurements indeed reveal a nearly balanced distribution of upward and downward $M_z$ domains. The histogram plot in Fig. 2g shows that $B_z$ follows a Gaussian distribution centered at ~0 G, suggesting the weak OOP magnetization $M_z$ (considering the polarity) is spatially averaged to be about zero over the sample area surveyed. When the MnTe sample is field cooled with $B_{FC} = -1$ T, a mostly negative stray field pattern emerges in both boundary and interior regions (Fig. 2b). The tunability by external field can be understood as follows. During magnetic field cooling, the Zeeman coupling between the OOP cooling field ($B_{FC}$) and the net magnetization lifts the degeneracy of the six in-plane easy axes along $[1\bar{1}00]$ to select the degenerate triplet state whose $M_z$ is parallel to $B_{FC}$, while the remaining three orientations are depopulated (Fig. 2e). Once the external magnetic cooling field is removed, the six domains become degenerate again, but relaxation from one triplet to the other requires the *Néel* order to rotate through the in-plane hard axis $[2\bar{1}\bar{1}0]$. The corresponding energy barrier suppresses the domain relaxation and leads to freezing of the field-cooled domain population. The histograms of normalized field percentages also corroborate this point (Fig. 2h). Lastly, when $B_{FC} = +1$ T, $M_z$ is spontaneously trained to upward direction, and it naturally favors the other three in-plane *Néel*-vector orientations (Fig. 2f).

The measured magnetic stray field map of the MnTe sample switches the polarity accordingly, as evidenced in Figs. 2h,i. It is worth mentioning that our experimentally measured stray field pattern agrees well with the simulated one with an assumption of uniform magnetization along the $z$-axis [67,70] (see Supplemental Material Section I.4 [51] for details). Certain minor regions with opposite field polarity observed during the magnetic field cooling processes are potentially due to domain walls formed between degenerate altermagnetic domains [19]. Under +1 T (−1 T) field cooling along the OOP direction, it is worth noting that $B_z$ measured over the boundary area is dominated by positive (negative) stray field with negligible opposite field, indicating a more uniform domain configuration compared to central regions, which is likely due to magnetostriction-induced formation of single altermagnetic domains [19]. We have performed extended scanning NV measurements on a hollowed hexagon device to further investigate this point (see Supplemental Material Section I.5 [51] for details). We also study MnTe magnetic domains in response to in-plane magnetic field cooling. The in-plane field cooling state shows a multidomain structure similar to the zero-field-cooling case (see Supplemental Material Section I.6 [51] for details), corroborating that the net magnetization in the MnTe film is oriented along the $c$-axis direction, consistent with the theoretical prediction.

Next, we report systematic scanning NV imaging of MnTe films with different thicknesses ($t$) to probe the origin of the weak magnetization. In antiferromagnets, uncompensated net magnetization can result from slight canting of the antiparallel magnetic moments, which is usually enhanced at the sample surface with a broken symmetry [71-76]. In this case, the weak magnetization is localized at the surface area, and the magnitude of two-dimensional (2D) magnetization $M_z^{2D}$, defined as the OOP magnetic moment normalized by lateral sample area, in principle, is insensitive to the sample thickness. On the other hand, if the nonzero $M_z$ in MnTe is bulk dominated, $M_z^{2D}$ as well as the resulting stray field are expected to increase linearly with film thickness. Figure 3a presents a series of stray field maps of MnTe samples with a thickness $t$ = 2 UC, 40 UC, 80 UC, and 230 UC measured at 2 K without field cooling. All the 4 MnTe samples are grown on InP(111)A substrates and are capped with 10-nm-thick Te. Multidomain features are observed in all samples, including the 2-UC-thick one, demonstrating long-range magnetic order established in atomically thin MnTe. The magnetic domain size increases monotonically with the sample thickness. A thicker MnTe film (230 UC) exhibits enhanced magnetic uniformity with larger magnetic domains, possibly due to reduced local defects, inhomogeneities, and/or an increase in grain sizes (see Supplemental Material Section I.7 [51] for details).

We observe that the measured magnetic stray field largely stays in the range from −1.5 G to +1.5 G as the sample thickness increases from 2 UC to 230 UC. To elucidate the origin of the observed MnTe magnetization, we reconstruct the magnetic moment [41,68] and normalize it by sample volume and lateral surface area, respectively (Figs. 3b-c). The reconstructed magnetization $M_z^{3D}$ (normalized by the sample volume) decreases rapidly as $t$ increases (Fig. 3b). In contrast, robust magnetic patches with $M_z^{2D}$ of a similar magnitude but opposite polarity are spatially distributed over MnTe films due to the absence of magnetic field cooling (Fig. 3c). To characterize the magnitude of $M_z^{3D}$ and $M_z^{2D}$, we calculate the corresponding root mean square (RMS) values $M_{z,RMS}^{3D}$ and $M_{z,RMS}^{2D}$ of MnTe films, as summarized in Fig. 3d. $M_{z,RMS}^{2D}$ is obtained to be 1.50 $\mu_B$/nm$^2$, 1.99 $\mu_B$/nm$^2$, 1.71 $\mu_B$/nm$^2$, and 1.98 $\mu_B$/nm$^2$ for the 2-UC-, 40-UC-, 80-UC- and 230-UC-thick MnTe films, whereas the $M_{z,RMS}^{3D}$ is 0.057 $\mu_B$/Mn, 0.004 $\mu_B$/Mn, 0.002 $\mu_B$/Mn, and 0.001 $\mu_B$/Mn, respectively. Despite the large variations of sample thickness by orders of magnitude, $M_{z,RMS}^{2D}$ shows a weak dependence on $t$ and follows a nearly constant value, highlighting that the weak net magnetization in MnTe films is surface dominated. It is instructive to note that our experimental value is about one order of magnitude larger than the one estimated from a bulk crystal or bulk theoretical model [59,61,65], also suggesting the surface instead of bulk origin of the net magnetization.

## C. Theoretical analysis

We now present the first-principles calculations to assess the relative magnetic contributions from the near-surface region versus the bulk interior in MnTe films. Here, we construct a slab model of a Te-capped MnTe film on an InP(111)A substrate (top of Fig. 4a), where the Te capping layer is modeled as an additional Te layer on the MnTe top surface. We compute the OOP magnetic moment in bulk MnTe, $m_{\mathrm{bulk}}$, and plot the atomic-layer resolved normalized $m_z/m_{\mathrm{bulk}}$ (Fig. 4a). Notably, the Te moment $m_{\mathrm{Te}}$ exhibits a weak atomic-layer dependence. In contrast, the Mn moment $m_{\mathrm{Mn}}$ is strongly enhanced by orders of magnitude at the MnTe/Te interface [69]. In the bulk, adjacent Mn layers are related by the mirror operation $\mathcal{M}_z$ such that the OOP magnetization $M_z$ on each Mn atom is identical despite their opposite in-plane components. At the (0001) surface, the $\mathcal{M}_z$ mirror symmetry is broken, so that the OOP magnetization $M_z$ is no longer symmetry-constrained to match its bulk value and is substantially enhanced in the near-surface region. We further calculate the orbital magnetic moments on Mn and Te atoms and find that the dominant contribution arises from the Mn spin moment, which exhibits the strongest enhancement at the surface and sets the overall magnitude of the local magnetization (See Supplemental Material Section I.8 [51] for details). In contrast, the Mn orbital moment is substantially smaller across all layers and shows only a weak layer dependence. The Te moments (both spin and orbital) remain close to zero throughout the slab with only minor variations near the surface, indicating that Te plays a secondary role in carrying the net magnetization. It is instructive to note that this defect-free calculation underestimates the measured $M_z^{2D}$ by roughly one order of magnitude. Motivated by the slightly $p$-type character of MnTe [61], we further calculate a slab containing a Mn vacancy $V_{\mathrm{Mn}}$ at the Te/MnTe interface, a defect expected to act as an acceptor and introduce hole carriers, as well as an alternative Te-capping configuration. Both calculations yield a surface-localized spin-dominated OOP moment, and the vacancy-containing model brings the calculated $M_z^{2D}$ in close agreement with our experimental value. Further discussion of our first-principles calculations on the surface magnetization is provided in Supplemental Material Section I.8 [51].

Below we invoke a symmetry argument to propose that the surface magnetization $M_z$ is uniquely determined by the bulk altermagnetic order in MnTe [62,69]. First, the surface magnetization $M_z$ remains invariant for the crystal configurations with the same bulk altermagnetic order but different surface terminations (Figs. 4b,c). Crystal configurations shown in Figs. 4b,c have the same bulk altermagnetic order, indicated by the dashed box, but their surfaces are terminated with different Mn sublattices. The two crystal configurations are related by two-fold rotation $C_{2z}$ along $z$-axis, which preserves the net surface magnetization $M_z$. Moreover, if the bulk altermagnetic order is reversed (Figs. 4c,d), the two configurations are related by the $C_{2z}T$ operation, under which surface magnetization $M_z$ is reversed. Here, $T$ is the time-reversal operation. Lastly, when considering multiple domains due to the hexagonal geometry, six in-plane easy axes of the *Néel* vector $\boldsymbol{L}$ can be classified into two symmetry-related triplets (Fig. 4e). Within each triplet, the three orientations separated by 120° are connected by the 3-fold rotation operation $C_{3z}$, which leaves the surface magnetization $M_z$ invariant. Two sets of triplets are the time-reversal $T$ symmetry partners, in which the $M_z$ is opposite. Thus, the surface magnetization of MnTe films is imprinted by their bulk altermagnetic order.

## D. Correlation with the AH effect in MnTe

Lastly, we establish the correlation between the observed net magnetization and the AH effect in MnTe films. In contrast to the conventional ferromagnetic materials where AH effect typically requires a robust macroscopic magnetization, altermagnets can exhibit a sizeable AH effect despite of a vanishingly small net magnetization [13,14]. This is because the AH effect in altermagnets is governed by the intrinsic Berry curvature of the electronic band structure instead of an extrinsic mechanism such as skew scattering or side jump that is driven by macroscopic magnetization [1,13]. Figure 5a presents scanning-probe NV imaging of a 40-UC-thick MnTe sample with $B_{FC}$ = 0 from 2 K to 300 K. The sample is grown on an InP(111)A substrate and is capped with 10-nm-thick Te. As temperature increases, the emanating stray field becomes weaker and eventually negligibly small within the signal-to-noise ratio above 250 K. We also perform AH measurements on the Te-capped 40-UC-thick MnTe sample from 200 K to 300 K. The linear contribution from the ordinary Hall effect has been subtracted in Fig. 5b. Clear AH signature is observed in the MnTe sample in the low-temperature regime. The coercive field and zero-magnetic-field AH resistance monotonically decrease with increasing temperature. The AH feature eventually vanishes when $T > 250$ K as evidenced by the nearly flat hysteresis loops, in agreement with our NV stray-field measurement results. To quantify the stray field signal, we use $\overline{B_P}$ and $\overline{B_N}$ to characterize average values of positive and negative stray fields generated by the upward and downward $M_z$ domains of the MnTe sample (Fig. 5c). The onset temperature $T_c$ for the OOP net magnetization $M_z$ is estimated to be ~250 K, consistent with that of the AH effect as shown in Fig. 5d. This correspondence underlines a strong correlation between the observed weak net magnetization and AH effect in MnTe films. To examine the dependence on the capping layer, we have prepared a separate MnTe(40 UC)/InP sample with a 2-nm-thick Bi capping, which also exhibits a comparable onset temperature (~250 K) for the net magnetization and AH effect (see Supplemental Material Section I.9 [51] for details). We further note that a slightly higher onset temperature and smaller domain size are observed in a 40-UC-thick MnTe film grown on a GaAs(111)B substrate (see Supplemental Material Section I.9 [51] for details), which is potentially attributed to lattice mismatch-induced epitaxial strain.

## III. Conclusion

In summary, we have utilized scanning-probe NV microscopy to spatially resolve the weak OOP net magnetization formed in epitaxial MnTe films down to the atomic thickness regime. We show that the nanoscale domain population associated with the underlying altermagnetic order can be controlled by an OOP magnetic cooling field. The observed MnTe magnetization exhibits a weak dependence on the film thickness, suggesting the surface-dominated contribution arising from spontaneous symmetry breaking. The onset of weak magnetization coincides with the appearance of the AH effect, revealing the underlying correlation between microscopic spin structure and magneto-transport response in altermagnetic MnTe films. Our results present a valuable ingredient of the emerging research topic of altermagnetism, bringing insights into future development of altermagnet-based spintronic devices. The current work also highlights the opportunities offered by quantum spin sensors in probing nanoscale electromagnetic properties of altermagnets, advancing the understanding of exotic spin behaviors in a broad family of magnetic quantum materials.

**Data availability**. All data supporting the findings of this study are available from the corresponding authors on reasonable request.

**Acknowledgments:** The authors are grateful to Elton Santos, Igor Mazin, and Kirill Belashchenko for helpful discussions. The quantum sensing measurements are primarily supported by the U.S. Department of Energy (DOE), Office of Science, Basic Energy Sciences (BES), under award No. DE-SC0024870. Development of cryogenic quantum microscopy is based upon work supported by the Air Force Office of Scientific Research under award No. FA9550-25-1-0082. The transport measurements are supported by the Office of Naval Research (ONR) under grant No. N00014-23-1-2146. Device fabrications are supported by U.S. National Science Foundation under awards No. DMR-2437294, No. ECCS-2445826 and No. ECCS-2525800. The MBE growth and XRD characterization are supported by the ONR Award (N000142412133). The ARPES measurements and theoretical calculations are supported by the seed project of the Penn State MRSEC for Nanoscale Science (DMR-2011839). C.-Z. C. acknowledges the support from the Gordon and Betty Moore Foundation's EPiQS Initiative (GBMF9063 to C.-Z. C). B. Y. acknowledges the financial support by the Israel Science Foundation (ISF: 2974/23).

**Author contributions**: L.-J. Z. performed quantum sensing measurements, transport measurements, and analyzed the data with the help of S. L. and H. L. Z.-J. Y., H. R., P. X., L.-K. L. performed the MBE growth, XRD characterization, and ARPES measurements under the supervision of C.-Z. C. Y. Z., H. B., C.-X. L., and B. Y. provided theoretical support and first-principles calculations. L.-J. Z., Z. X., Y. Z., and Z. G. fabricated the photolithographically patterned devices. K. W. performed the STEM measurements. H. W. contributed to the discussion and data interpretation. C. D. supervised this project.

**Competing interests**

The authors declare no competing interests.

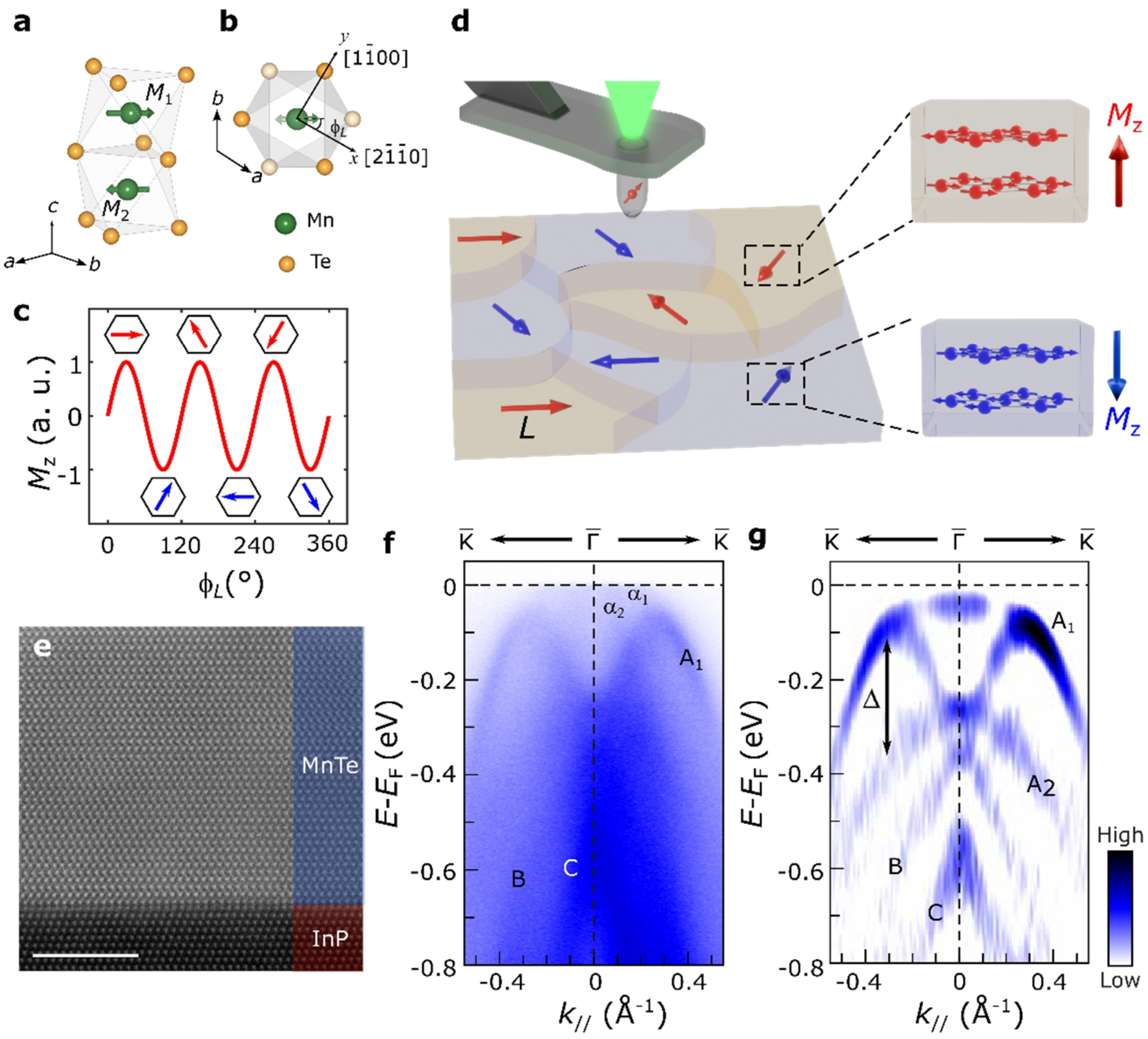

**Fig. 1. Crystal structure and characterizations of epitaxial MnTe films.** (a) Schematic of the crystal structure and antiferromagnetic order of MnTe. The *Néel* vector $\boldsymbol{L} = \boldsymbol{M}_1 - \boldsymbol{M}_2$ is defined by two antiparallel magnetic moments ($\boldsymbol{M}_1$ and $\boldsymbol{M}_2$) of Mn atoms on neighboring sublattices. (b) $\phi_L$ denotes the angle between the *Néel* vector $\boldsymbol{L}$ and the $[2\bar{1}\bar{1}0]$ crystallographic axis in the *c*-plane of MnTe. (c) Theoretical prediction of the weak OOP magnetization $M_z$ as a function of $\phi_L$. Two groups of red and blue arrows represent the six in-plane easy axes of the *Néel* vector $\boldsymbol{L}$, classified into two symmetry-related triplets. (d) Schematics of scanning-probe NV imaging of local magnetic stray fields generated by the net magnetization $M_z$ in an MnTe film. Domains with two different *Néel* vector $\boldsymbol{L}$ orientations and the corresponding $M_z$ configuration are highlighted on the right. (e) STEM image of a 40-UC-thick MnTe film grown on an InP(111)A substrate. The scale bar is 5 nm. (f) ARPES band dispersion of an 8-UC-thick MnTe film along the $\bar{K} - \bar{\Gamma} - \bar{K}$ direction measured at 2 K. Several bulk bands cross at the $\bar{\Gamma}$ point, located at ~0.1 eV (labeled as $A_1$), ~0.4 eV (labeled as B), and ~0.7 eV (labeled as C) below the Fermi level. The Fermi level crosses two surface states (labeled as $\alpha_1$ and $\alpha_2$). (g) Second-derivative band map with respect to the energy of the band structure. Spin-splitting $\Delta \sim 280$ meV below the Fermi level is resolved.

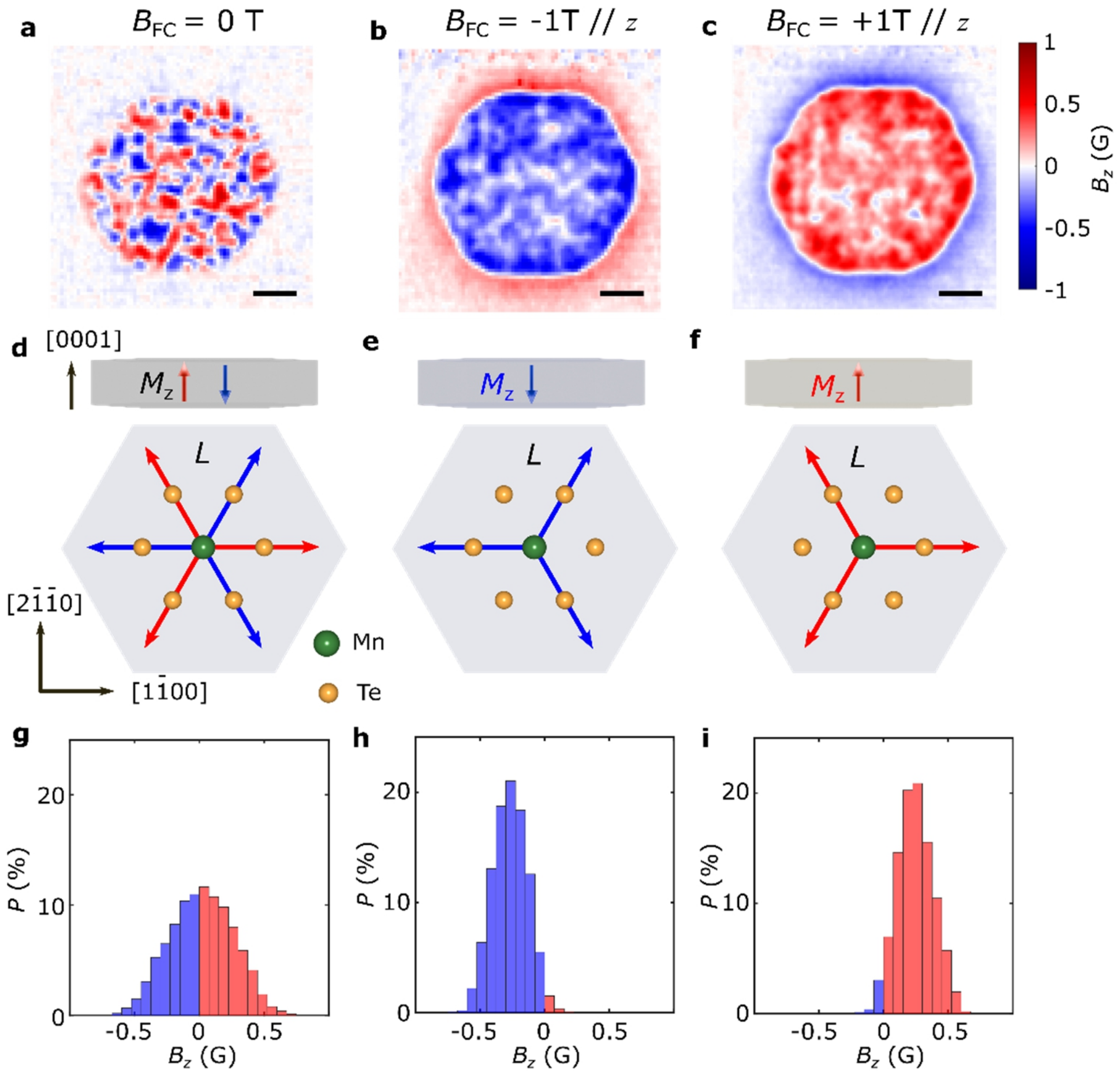

**Fig. 2. Scanning-probe NV imaging of field-cooling-controlled MnTe magnetic domains.** (a)-(c) Magnetic stray-field maps of a 40-UC-thick MnTe pattern (~10 μm diameter and grown on an InP substrate) measured under OOP-oriented magnetic cooling fields $B_{FC}$ = 0 (a), −1 T (b), and +1 T (c). The scale bar is 2 μm. (d)-(f) Schematics of the possible configurations of the in-plane *Néel* vector $\boldsymbol{L}$ and the corresponding $M_z$ in MnTe for $B_{FC}$ = 0 (d), −1 T (e), and +1 T (f). The six arrows within the hexagon denote the six easy axes of $\boldsymbol{L}$ (bottom), color-coded by their corresponding $M_z$ polarity (top). (g)-(i) Corresponding histograms of $B_z$ distributions extracted from the MnTe sample area of the NV imaging maps in (a)-(c). All scanning-probe NV magnetometry measurements are performed at $T$ = 2 K.

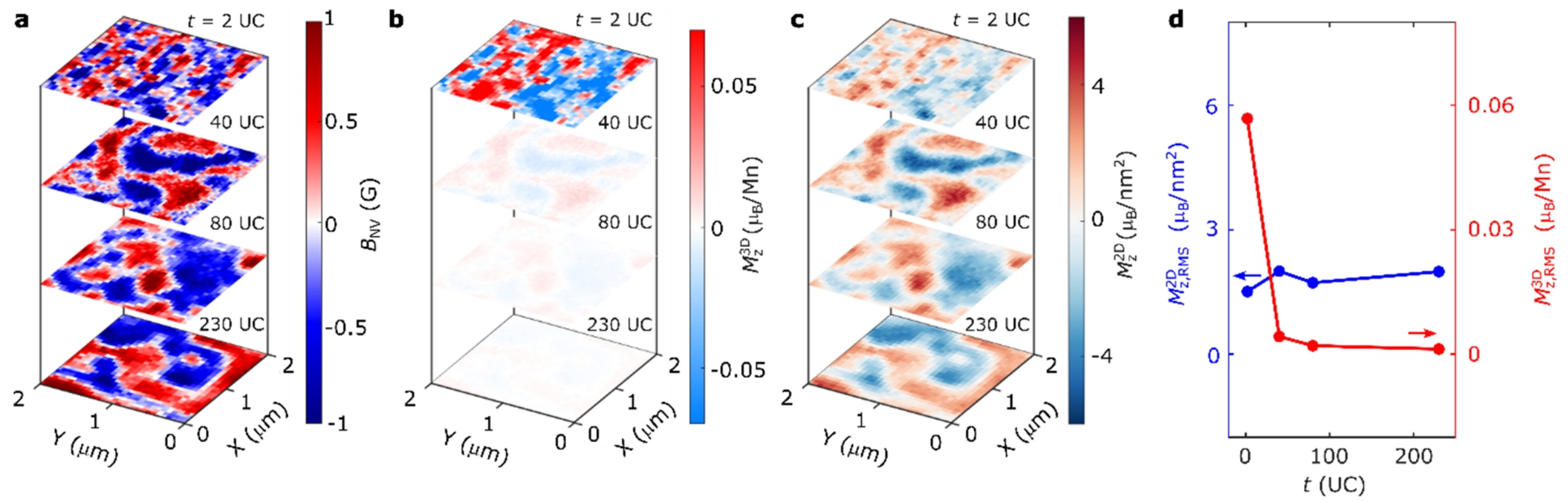

**Fig. 3. Surface-dominated origin of the weak net magnetization in MnTe films.** (a)-(c) Magnetic stray field $B_{\mathrm{NV}}$ along the NV axis (a), 3D magnetization $M_z^{3D}$ (b), and 2D magnetization $M_z^{2D}$ (c) maps for MnTe/InP samples with MnTe thicknesses of 2-UC, 40-UC, 80-UC, and 230-UC. All NV measurements are performed at 2 K. (d) Root-mean-square values of $M_z^{2D}$ (blue) and $M_z^{3D}$ (red) as a function of the MnTe film thickness $t$.

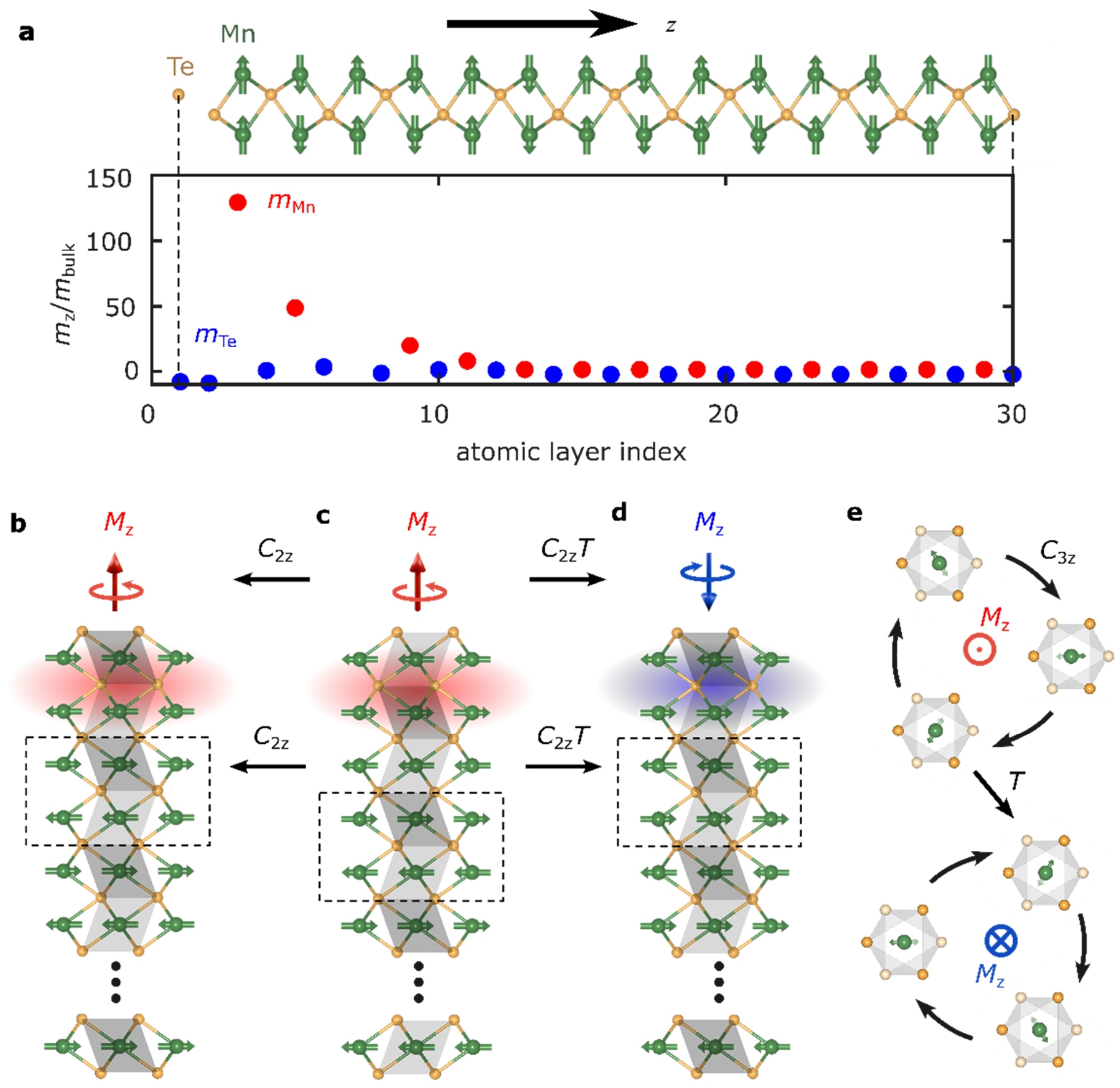

**Fig. 4. Surface magnetization and its correlation to the bulk altermagnetic order.** (a) Top: Schematic of the Te-capped MnTe slab. The atomic layer index starts from the surface of Te capping layer and increases towards the MnTe bulk. Bottom: First-principles calculations of layer-resolved magnetic moment $m_z$ for Mn (red) and Te (blue) atoms as a function of the atomic layer index. (b-d) The reference MnTe spin configurations (c) and its symmetry-related counterparts under $C_{2z}$ (b) and $C_{2z}T$ (d) operation. The bulk altermagnetic order is outlined by the dashed box. Two Mn sublattices with different Te environments are labeled by different shadow boxes. The top panel shows the net surface magnetization $M_z$ under symmetry operation. (e) Six possible domains form two triplets of *Néel* vectors separated by 120°. Within a triplet, three members are related by the $C_{3z}$, which preserves the sign of $M_z$ (red, out of the plane). The two triplets are related to each other by a $T$ operation, which reverses the sign of $M_z$ (blue, into the plane).

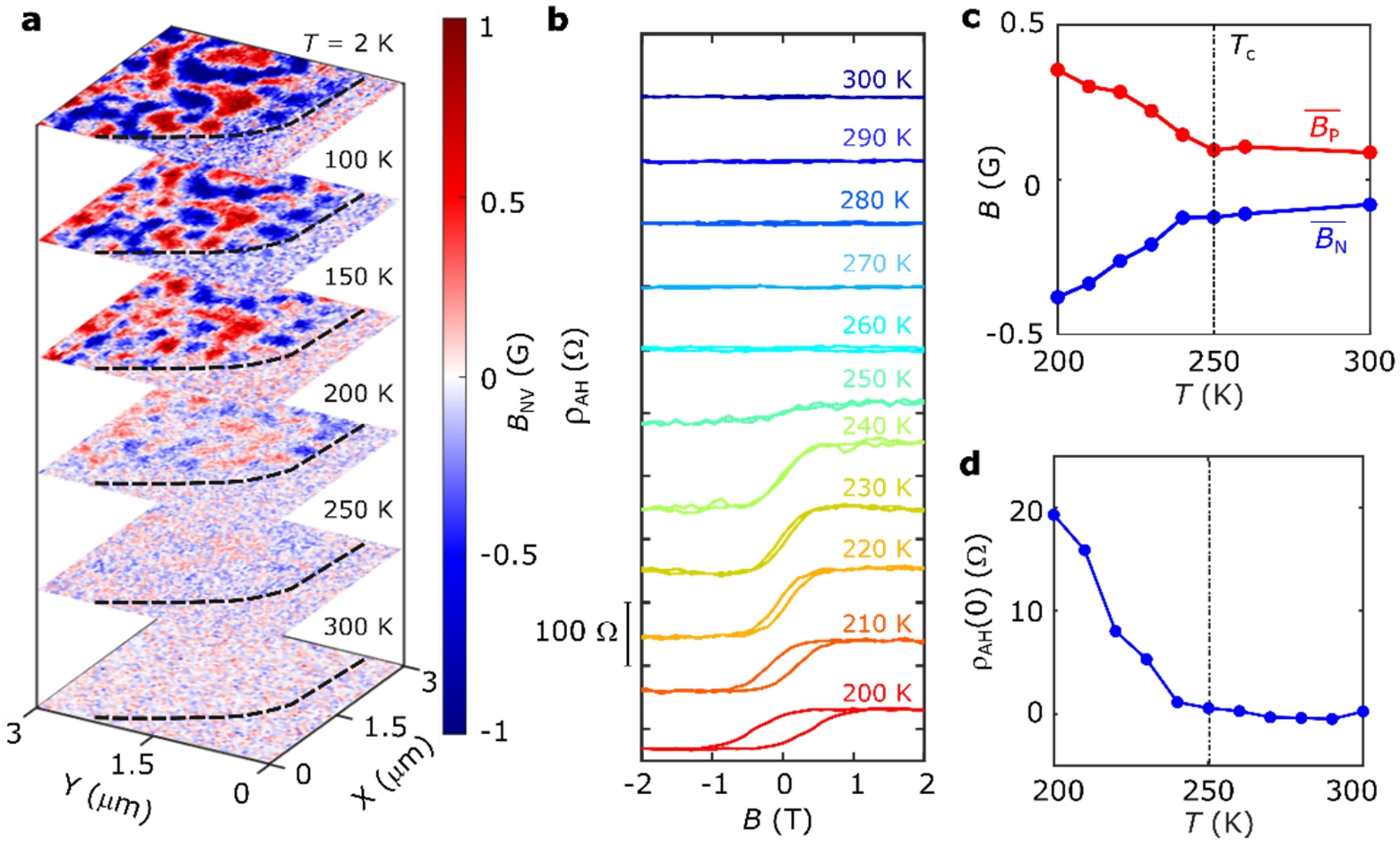

**Fig. 5. Correlation between the weak net magnetization and AH effect in MnTe films.** (a) Scanning-probe NV imaging of magnetic stray field $B_{NV}$ measured along NV axis from an MnTe(40 UC)/InP sample at $T$ = 2 K, 100 K, 150 K, 200 K, 250 K, and 300 K. Black dashed lines outline the boundary of the MnTe pattern. (b) AH resistance $\rho_{AH}$ of the same sample as a function of the OOP magnetic field measured from 200 K to 300 K. (c) $T$ dependence of the averaged positive and negative magnetic stray fields $\overline{B_P}$ (red) and $\overline{B_N}$ (blue), respectively, emanating from the MnTe sample. (d) Temperature ($T$) dependent zero-magnetic-field AH resistance $\rho_{AH}(0)$ of the MnTe sample.